# Aluminium storage using nitrogen-doped graphene nanoribbons from first principles

Jovana Vlahović[1], Ana S. Dobrota[1*], Natalia V. Skorodumova[2,3], Igor A. Pašti[1,2]

[1] *University of Belgrade – Faculty of Physical Chemistry, Belgrade, Serbia*
[2] *Department of Materials Science and Engineering, School of Industrial Engineering and Management, KTH-Royal Institute of Technology, Stockholm, Sweden*
[3] *Department of Physics and Astronomy, Uppsala University, Uppsala, Sweden*

## Abstract

Pristine graphene interacts relatively weakly with Al, which is a specie of importance for novel generations of metal-ion batteries. We employ DFT calculations to explore the possibility of enhancing Al interaction with graphene. We investigate non-doped and N-doped graphene nanoribbons, address the impact of the edge sites, which are always present to some extent in real samples, and N-containing defects on the material's reactivity towards Al. The results are compared to that of pristine graphene. We show that introduction of edges does not affect the energetics of Al adsorption significantly by itself. On the other hand, N-doping of graphene nanoribbons is found to affect the adsorption energy of Al to the extent that strongly depends on the type of N-containing defect. While graphitic and pyrrolic N induce minimal changes, the introduction of edge NO group and doping with in-plane pyridinic N result in Al adsorption nearly twice as strong as on pristine graphene. The obtained results could guide the further design of advanced materials for Al-ion rechargeable batteries.

## Keywords

Graphene; Nanoribbons; Aluminium; Adsorption; DFT; Metal-ion battery.

*corresponding author, e-mail: ana.dobrota@ffh.bg.ac.rs

# 1. Introduction

Graphene, a 2D honeycomb structure of sp$^2$ hybridized C atoms, is assumed to be a promising material for various applications due to low mass, two-dimensionality, excellent electrical conductivity and other fascinating properties [1]. However, due to its chemical inertness (in the pristine state), it is often necessary to functionalize or dope it to make it suitable for a chosen purpose. This is especially true when it comes to energy-related applications.

The world is energy-addicted, and the energy needs are ever rising. Among the novel energy conversion and storage systems, Al-ion batteries stand out due to their low cost and low flammability. Small mass and radius, and high abundance of Al in Earth's crust, together with three-electron redox properties, lead to a high capacity of Al-ion batteries [2]. Carbon-based materials are widely investigated as candidates for electrode materials in Al-ion batteries. Lin *et al.* have reported a secondary aluminium-ion battery with an aluminium metal anode, a graphitic-foam cathode and an ionic liquid electrolyte containing chlorides, operating through chloroaluminate anions' (de)intercalation between graphite layers during battery (dis)charge [3]. Jiao *et al.* have reported a rechargeable Al-ion battery that uses ionic liquid electrolyte containing chlorides, high-purity Al foil anode, and carbon paper cathode, which can be completely recycled to obtain graphene [4]. Graphene is assumed to be a good electrode material due to its properties mentioned above, but Al interacts relatively weakly with pristine graphene [5], resulting in low battery voltage. In that sense, it is essential to find a way to enhance Al adsorption onto graphene-based materials. The chemical inertness of pristine graphene can be overcome by doping or functionalization. The most investigated dopant atoms are B and N, due to similar atomic size as C, which do not perturb the planarity of graphene. Numerous studies report that doping graphene by nitrogen can enhance its catalytic ability for energy conversion and storage applications, but the mechanism of this enhancement is still unclear [6]. Rao *et al.* [7] provided a comprehensive review on the synthesis, characterization, and application of nitrogen and boron-doped graphene. Various studies have shown that nitrogen-doped graphene is very suitable for energy storage and conversion systems [8,9]. It was also shown that the adsorption of lithium is enhanced on nitrogen-doped graphene and that it is suitable for Li-ion battery applications [9]. In our previous work, we showed that the presence of the oxygen functional groups on the graphene plane can strengthen the adsorption of alkali metals [10], as well as that the presence of some dopants can enhance Na adsorption on graphene [11].

Besides 2D graphene, other prospective carbon allotropes are considered for various applications: one-dimensional nanotubes, quasi-one-dimensional nanoribbons, and zero-dimensional fullerenes. The performance of these nanomaterials depends on their structure, physical and chemical properties. For example, while pristine graphene is a zero-gap semimetal, the resistivity of a graphene nanoribbon increases as its width decreases [12], and

sub-10-nanometer graphene nanoribbons are semiconductors [13]. An interesting strategy for obtaining targeted material properties is combining the dimensionality of carbon with doping and functionalization. Doping graphene nanowires with B and N was investigated in ref. [14]. It was concluded that graphene N-doped nanowires have a more favorable distribution of energy levels and that the Fermi level is lower than in the case of doping with boron. Obviously, nitrogen-doped graphene strips would be an interesting candidate for use in energy storage and conversion devices. The impact of a dopant atom on the electronic and chemical properties of graphene crucially depends on the bonding environment (configuration) of the guest atom [15]. For example, in N-doped graphene, the presence of pyridinic nitrogen results in p-type doping, while graphitic N donates excess electrons to the conduction band and converts graphene to an n-type semiconductor [15]. Obviously, it is of utmost importance to control precisely the bonding configuration of defects embedded in graphene. In the literature [14], the influence of the position of nitrogen as a dopant in the structure of the graphene nanoribbon was investigated. It is shown that the Fermi level is lowest when nitrogen is at the edge of the nanoribbon, and then the material behaves like an n-type semiconductor. Ref. [16] also shows that doping at the edge is more energy-efficient than doping at sites in the centre of the nanoribbon. In the same report, sodium adsorption on N-doped graphene nanoribbon at different sites was tested. The authors have shown that the adsorption is the strongest near the dopant located at the very edge of the graphene nanoribbon.

In order to investigate aluminium interaction with several graphene-based surfaces, we performed Density Functional Theory (DFT) calculations with the idea of their possible use as electrode material in aluminium-ion batteries. We study the adsorption of aluminium on pristine graphene and graphene nanoribbons saturated with hydrogen atoms on the edges (GNRH) and GNRH with various nitrogen defects. We analyse changes in the electronic structure of these systems and try to link them to their chemical properties.

## 2. Computational details

The DFT calculations were performed using the Quantum ESPRESSO package [17,18], for *ab initio* calculations. Used PWscf code has incorporated ultra-soft pseudopotentials and generalized gradient approximation with the Perdew-Burke-Ernzerhof (GGA PBE) exchange-correlation functional [19]. In all the calculations, the plane waves kinetic energy cut-off was 36 Ry, while the charge density cut-off was 576 Ry. All calculations were spin-polarized. Atomic positions were fully relaxed until the residual forces acting on atoms were below 0.001 eV Å$^{-1}$.

We use pristine graphene (*p*-graphene) as the benchmark and the simplest model in our periodic calculations. We use a hexagonal graphene supercell consisting of 54 carbon atoms ($C_{54}$), corresponding to the ($3\sqrt{3}\times3\sqrt{3}$)$R$30° structure. The two-dimensionality of the system is achieved by the sufficiently large height of the supercell (24 Å), so that the adjacent layers of graphene (along the $z$-axis) do not interact. The first irreducible Brillouin zone was sampled using $\Gamma$-centred 4×4×1 set of $k$-points generated using the general Monkhorst-Pack scheme [20]. Then, a denser, 20×20×1 set of $k$-points was used to analyse the densities of electronic states (DOS).

The models of (nitrogen-doped) graphene nanoribbons used in this contribution have been reported in our previous work, ref. [21]. Therefore, we will describe them here only briefly. The simplest graphene nanoribbon was derived from $p$-graphene by saturating C atoms on the supercell edges parallel to its $y$-axis by hydrogen atoms and by increasing the $x$-parameter of the supercell. That way, we obtain the simplest graphene nanoribbon saturated with hydrogen (GNRH), $C_{54}H_{12}$. The elongation of the supercell along the $x$-axis ensures no interactions between adjacent nanoribbons in the periodic calculation. This was also ensured by using a $\Gamma$-centred 1×4×1 set of $k$-points generated using the general Monkhorst-Pack scheme [20] to represent the first two-dimensional Brillouin zone. A denser, 1×32×1 set of $k$-points was used for DOS analysis. N-containing nanoribbons were obtained by modifying the $C_{54}H_{12}$ model by adding one nitrogen functional group into the supercell, resulting in N concentration of approximately 1.5 at.%. That way we obtain: (i) GNRH in which one C atom is substituted with one N atom (graphitic N with three sp$^2$ C−N bonds, model $C_{53}NH_{12}$), (ii) GNRH doped by pyridinic N at the edge (two C−N bonds in a hexagon, model $C_{53}NH_{11}$), (iii) GNRH doped by pyridinic N inside the basal plane, saturated by H (model $C_{52}NH_{14}$: two C−N bonds in a hexagon inside the plane, with one in-plane C-monovacancy, two dangling C atoms saturated by H), (iv) GNRH doped by pyridinic N inside the basal plane, unsaturated by H (model $C_{52}NH_{12}$), (v) GNRH doped by pyrrolic N at the edge (so that one former edge $C_6$ hexagon becomes a $C_4N$ pentagon) saturated by H ($C_{52}NH_{11}$ model); (vi) GNRH doped by N-oxide at the edge ($C_{53}H_{11}NO$), and (vii) GNRH functionalized by an amino group at the edge site, instead of one H ($C_{54}H_{11}NH_2$). Whereas in ref. [21] we reported four variations of $C_{53}NH_{12}$ with different positions of quaternary N, here we use only the most stable one of those four models. For graphical presentation of all the GNRH models please see ref. [21].

During the interaction of many elements of the periodic table with graphene dispersion interactions significantly contribute to the overall interaction [5]. That is also the case for Al on graphene [5]. Therefore, in this contribution, all the calculations are done with the dispersion correction, using Grimme's DFT-D2 scheme [22] implemented in Quantum ESPRESSO [17,18]. Dispersion interactions are pair-type, calculated for each pair of atoms in the system and added to the total energy. The strength of the interaction of Al with the chosen substrates is expressed through its adsorption energy. Adsorption energies of Al ($E_{ads}$(Al)) are evaluated as:

$$E_{ads}(Al) = E_{sub+Al} - (E_{sub} + E_{Al}) \tag{1}$$

where $E_{sub+Al}$, $E_{sub}$, and $E_{Al}$ stand for ground-state total energies of the substrate with Al adsorbed, the total energy of a bare substrate, and the total energy of an isolated Al atom, respectively. As defined, negative $E_{ads}$ indicates that an exothermic process has taken place. For charge analysis, we used Bader's algorithm [23] on a charge density grid by Henkelman *et al*. [24]. For 3D charge difference plots, charge difference ($\Delta\rho$) was calculated as:

$$\Delta\rho = \rho_{Al@subs} - (\rho_{Al@subs-Al} + \rho_{Al@subs-subs}) \tag{2}$$

where $\rho_{Al@subs}$ stands for the ground-state charge density of the substrate with Al adsorbed in the optimal position (as given in **Fig. 2**), $\rho_{Al@subs-Al}$ the ground-state charge density of the system with the same geometry as the previous one, only without Al atom, and $\rho_{Al@subs-subs}$ the ground-state charge density of isolated Al.

## 3. Results and Discussion
### 3.1. Al adsorption on pristine graphene

There are three high-symmetry sites available for any adsorbate on pristine graphene: (i) above one C atom (top site), (ii) above the centre of a C−C bond (bridge site), and (iii) above the centre of the $C_6$ hexagon (hollow site). With the presence of defects in graphene's structure, the number of inequivalent adsorption sites increases. We find that the preferential site for Al adsorption on *p*-graphene is the hollow site (**Fig. 1 (a)** and **(b)**), where Al is hovering ≈ 2.09 Å above the basal plane, with the adsorption energy of −1.25 eV. This result agrees with available literature data, where the hollow site is also identified as the most favourable one for Al on graphene, with similar adsorption energies (**Table 1**), depending on the computational approach and supercell size (aluminium concentration). We find that the inclusion of dispersion interactions does not change Al's site preference (hollow site is preferred both with and without the dispersion correction), as can also be seen from ref. [5], except for the case of the non-local vdW-DF2 functional (**Table 1**), which we do not use here.

The significantly stronger Al binding reported in ref. [26], compared to other available data and this study, can be ascribed to the fact that Local Density Approximation (LDA) tends to result in over-bonding in molecular and solid systems. However, it is used nevertheless to mimic the effect of van der Waals interactions (but with the wrong physical background) [28] instead of treating it explicitly.

**Table 1.** Al adsorption on pristine graphene: available literature data and the results of this study. For each case, the used simulation package, computational scheme, graphene model (supercell size), calculated Al adsorption energy ($E_{ads}$(Al)) and Al height above the graphene layer ($h$) are given.

| Simulation package | Computational scheme | Graphene model | Preferred Al ads. site | $E_{ads}$(Al) / eV | $h$ / Å | Source |
|---|---|---|---|---|---|---|
| VASP | GGA PBE | $C_{32}$ (4×4) | hollow | −1.042 | 2.13 | ref. [25] |
| VASP | LDA[a] | $C_{18}$ (3×3) | hollow | −1.62 | 2.04 | ref. [26] |
| VASP | GGA PBE | $C_{32}$ (4×4) | hollow | −0.88 | 2.14 | ref. [5] |
| | GGA PBE-D2[b] | | hollow | −1.09 | 2.14 | |
| | GGA PBE-D3[c] | | | −0.99 | 2.19 | |
| | GGA vdW-DF2[d] | | bridge | −0.60 | 2.00 | |
| VASP | GGA PBE | $C_{32}$ (4×4) | hollow | −0.77 | 2.12 | ref. [27] |
| | GGA PBE-D2 | | | −0.98 | 2.11 | |
| | GGA PBE-D3 | | | −0.88 | 2.19 | |
| | GGA vdW-DF2 | | | −0.44 | 2.31 | |
| Quantum ESPRESSO | GGA PBE | $C_{54}$ ((3√3×3√3)$R$30°) | hollow | −1.00 | 2.11 | this study |
| | GGA PBE-D2 | | | −1.25 | 2.09 | |

[a] Local Density Approximation; [b] GGA PBE with Grimme's D2 dispersion correction; [c] GGA PBE with Grimme's D3 dispersion correction; [d] non-local vdW-DF2 functional.

The differences between the results reported here and the other literature data shown in **Table 1** are mostly a consequence of varying graphene supercell size (*i.e.*, differing Al concentrations). Same as for the case of Li adsorption on *p*-graphene [29], Al binding is stronger for lower metal concentrations (**Table 1**, GGA PBE-D2 scheme) since charge transfer is largest for lowest metal concentrations [29]. Additionally, while in some of the mentioned studies all of the atoms were allowed to find their optimal positions during the relaxation of Al@*p*-graphene systems, in others the positions of certain atoms were fixed, resulting in somewhat altered Al adsorption energies. As was outlined by Fernandez *et al.*, this effect amounts to ≤5% of $E_{ads}$(Al) [27]. Moreover, spin polarization can alter $E_{ads}$(Al) by about 0.15 eV [27]. Studies dealing with Al adsorption on slab models of graphene are excluded from this discussion as we explore ideal, pristine graphene in this section. Considering everything stated above, we consider our choice of computational method (section 2) well thought out.

Looking at DOS plots of Al@$C_{54}$ (**Fig. 1(c)**) we observe an upshift of the Fermi level indicating that graphene gains charge upon Al adsorption. According to Bader charges approx. 1.30 e is transferred from Al to graphene upon adsorption.

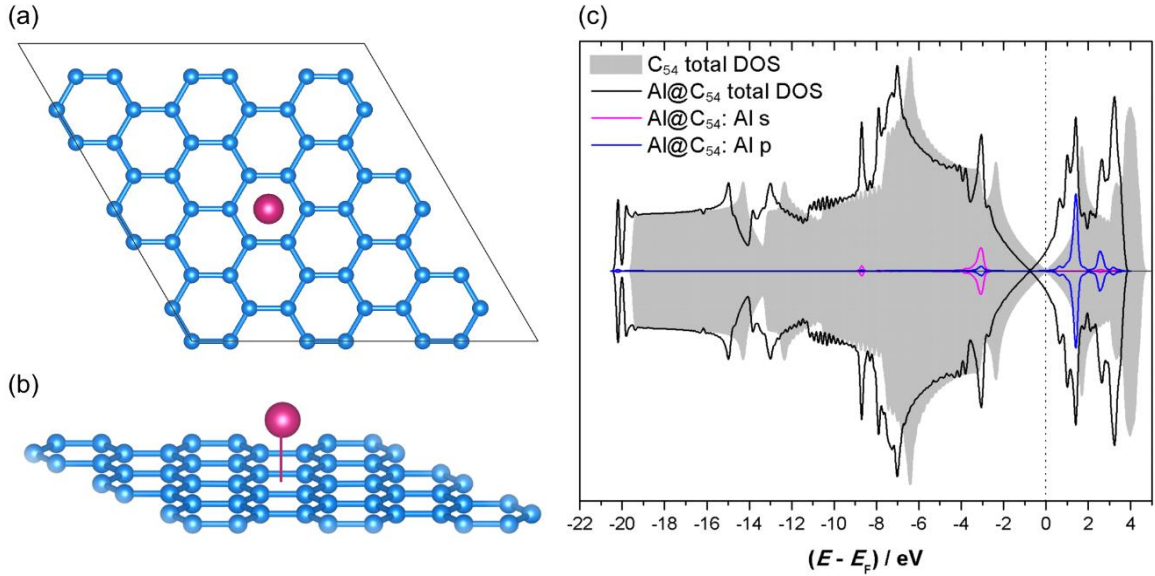

**Figure 1.** Al adsorption on the preferential site of *p*-graphene (hollow site). The optimized structure, made using VESTA [30], is shown from the top (a), including the supercell borders, and from the side (b). Total DOSes of pristine graphene before (grey shade) and upon Al adsorption (black line) are shown in (c), including Al's s and p states as well.

### 3.2. Al adsorption on (nitrogen-doped) graphene nanoribbons

The adsorption of Al is investigated on different non-equivalent sites of GNRH with and without N-containing defects. Generally, it could be expected that Al adsorption would be weakened by N doping of GNRH because N introduces an additional electron to the system, hindering charge transfer from Al to the substrate. But, of course, this strongly depends on the type of N defect, as will be discussed below.

In the case of Al adsorbed on $C_{54}H_{12}$, we find a situation similar to Al on *p*-graphene. The preferential adsorption site for Al@$C_{54}H_{12}$ is the modified edge hollow site (**Fig. 2(a)**), and the corresponding adsorption energy is −1.25 eV. By "modified edge hollow site" we describe the fact that the optimal Al position is not exactly above the $C_6$-ring centre, but rather shifted towards the HC−CH bridge at the edge (**Fig. 2(a)**). It should be noted that Al adsorption on other hollow sites of $C_{54}H_{12}$ results in adsorption energies which are more positive than on the preferential site by up to 0.06 eV, making the effect of the site's distance from the edge on $E_{ads}$(Al) rather weak and indicating high mobility of Al on this surface. Al adsorption on the preferential site of $C_{54}H_{12}$ induces some structural deformation to the substrate (**Fig. 2(a)**), as the plane gets distorted near the edge (two edge C atoms protrude the plane by approx. 0.3 Å, and H atoms bound to them by about 0.4 Å). It is interesting to note that $E_{ads}$(Al) is the same (to the 2$^{nd}$ decimal) on $C_{54}$ and $C_{54}H_{12}$ models, indicating that introducing the hydrogen saturated edges does not significantly affect Al adsorption energetics. However, in the case of Al@$C_{54}H_{12}$, Al transfers ≈ 0.78 e more to the substrate and is closer to it by about 0.17 Å, compared to Al@$C_{54}$.

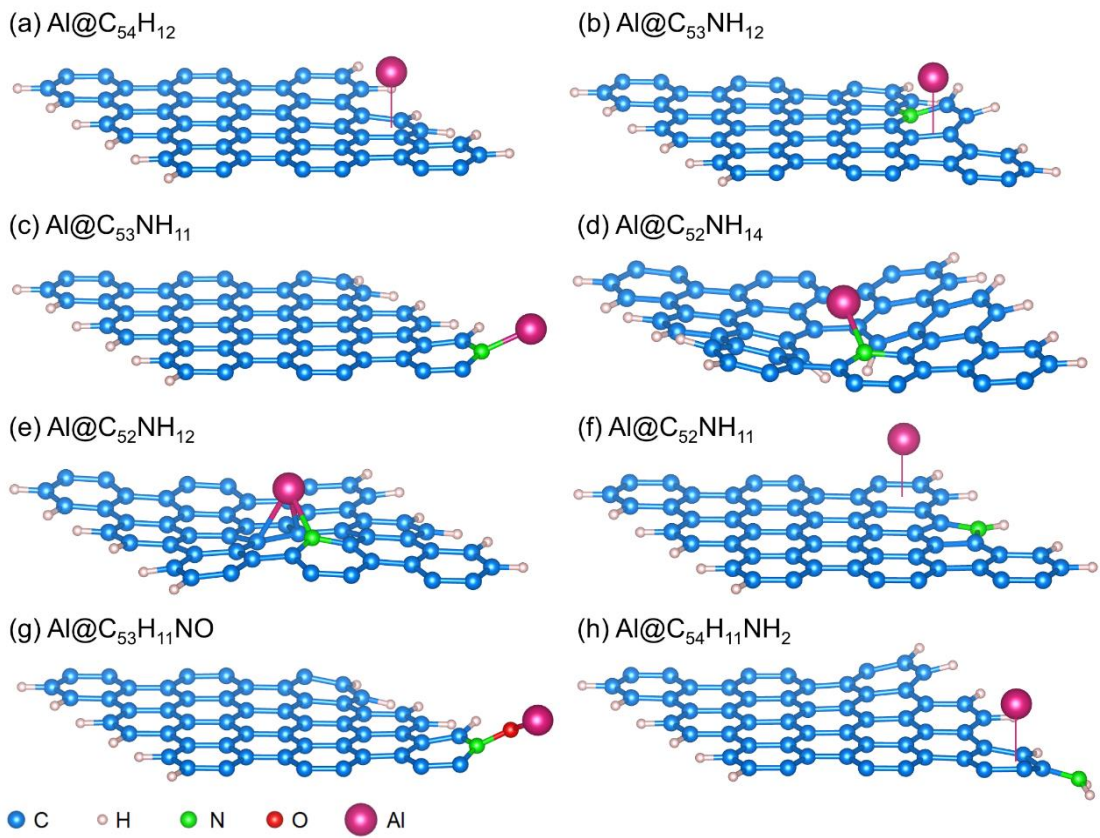

**Figure 2.** Al adsorption on the preferential sites of chosen GNRH: (a) non-doped $C_{54}H_{12}$, (b) graphitic N-doped $C_{53}NH_{12}$, (c) edge pyridinic N-doped $C_{53}NH_{11}$, (d) saturated in-plane pyridinic N-doped $C_{52}NH_{14}$, (e) unsaturated in-plane pyridinic N-doped $C_{52}NH_{12}$, (f) pyrrolic N-doped $C_{52}NH_{11}$, (g) N-oxide doped $C_{53}H_{11}NO$ and (h) amino-functionalized $C_{54}H_{11}NH_2$. Graphical representations were made using VESTA [30].

If the graphene nanoribbon is doped with graphitic N ($C_{53}NH_{12}$ model), Al adsorbs preferentially onto the C−C bridge site within the $C_5N$ ring, away from N, to a site that does not contain edge C atoms saturated with H (**Fig. 2(b)**). Al transfers significantly less charge to the substrate, compared to the previous case, **Table 2**, and the corresponding $E_{ads}(Al)$ is −1.21 eV. The substrate, which was planar (before Al adsorption), gets distorted at the edge, similar to the case of Al@$C_{54}H_{12}$, but with somewhat larger atoms protrusion (approx. 0.5 Å for edge C atoms and 0.8 Å for H atoms bound to them). Obviously, Al adsorption is slightly weakened (by 0.04 eV) by the presence of graphitic N. This agrees with the experimental findings of Lin *et al.,* who showed that metal ions do not tend to be trapped onto the graphitic N site in graphene [6]. Our findings suggest that that is also the case for graphitic N in GNRH, *i.e.*, the introduction of edges does not significantly affect the reactivity of quaternary N in graphene.

In the case of the nanoribbon containing pyridinic N at the edge ($C_{53}NH_{11}$), the strongest Al adsorption is observed when Al interacts directly with N (**Fig. 2(c)**). The corresponding $E_{ads}(Al)$ is −1.77 eV, and the charge is transferred from Al to $C_{53}NH_{11}$ (**Table 2**). If pyridinic

N is located inside the graphene plane ($C_{52}NH_{14}$), rather than on its edge, Al bonding is even stronger, with the corresponding $E_{ads}$(Al) of −2.74 eV (**Table 2**). In this case, Al preferentially interacts with N, near the C-vacancy site (**Fig. 2(d)**), transferring 1.34 e to the substrate, which is significantly corrugated. Obviously, the presence of pyridinic N, either on edge or inside the plane, enhances Al adsorption compared to all previous substrates. This coincides with the finding of the study, which showed that the presence of three and four adjoining pyridinic N sites in graphene provides a favourable site for trapping metal ions [6]. We emphasize that in our study, in the in-plane pyridinic N case, two C atoms with dangling bonds (due to the presence of one C-monovacancy, due to the presence of pyridinic N) were saturated by H. However, if they would not be saturated, the corresponding $E_{ads}$(Al) would be even more negative, −3.39 eV, when Al is adsorbed at the vacancy site (**Fig. 2(e)**). Compared to the case of a single vacancy in pristine graphene, this adsorption energy is more positive by 1.98 eV [31], indicating that the presence of N lowers the vacancy's reactivity. Due to the high reactivity of an unsaturated, C-vacancy-pyridinic-N-site, we consider it unlikely for it not to be saturated and find the saturated in-plane pyridinic model more realistic in everyday laboratory conditions. Unsaturated edges and vacancies are known to dominate under high-vacuum conditions and during non-equilibrium high-temperature growth [32].

The case of Al adsorption on GNRH doped with pyrrolic N at the edge ($C_{52}NH_{11}$ model) is somewhat similar to the case of graphitic N - the adsorption energy is 0.05 eV more negative than on $C_{53}NH_{12}$, and 0.08 e more is transferred from Al to the substrate (**Table 2**). However, the preferred adsorption site, in this case, is the edge hollow site which is the furthest from pyrrolic N (**Fig. 2(f)**). Evidently, graphitic and pyrrolic N are not very useful defects for enhancing the reactivity of GNRH towards Al.

When GNRH is doped with NO ($C_{53}H_{11}NO$ model), Al prefers direct interaction with O atom (**Fig. 2(g)**), with the adsorption energy of −2.43 eV and charge transfer from Al to the substrate (**Table 2**). This one results in the strongest Al bonding from all the investigated cases, nearly as twice as strong as on *p*-graphene. This was expected, since the NO group consists of two highly electronegative atoms willing to accept Al's charge. Upon Al adsorption onto $C_{53}H_{11}NO$, O obtains approx. 0.6 e additional charge, while the charge at N changes only slightly.

Functionalization of GNRH with the amino group ($C_{54}H_{11}NH_2$ model) results in preferential Al adsorption on the modified edge hollow/bridge site containing the carbon atom which is bound to $NH_2$ (**Fig. 2(h)**). The corresponding adsorption energy is −1.37 eV, while 1.22 e is transferred from Al to $C_{54}H_{11}NH_2$.

**Table 2.** Al adsorption on the preferential sites of non-doped and N-doped GNRH: corresponding adsorption energies ($E_{ads}$(Al)), charge transferred from Al to the substrate ($\Delta q$(Al)) and Al's distance from its nearest neighbour ($d$).

| Model | Dopant type | $E_{ads}$(Al) / eV | $\Delta q$(Al) / e | Al's nearest neighbour | $d$ / Å |
|---|---|---|---|---|---|
| $C_{54}H_{12}$ | non-doped | −1.25 | 2.08 | C | 2.44 |
| $C_{53}NH_{12}$ | graphitic N | −1.21 | 1.07 | C | 2.27 |
| $C_{53}NH_{11}$ | edge pyridinic | −1.77 | 1.03 | N | 1.96 |
| $C_{52}NH_{14}$ | in-plane pyridinic, saturated | −2.74 | 1.34 | N | 2.10 |
| $C_{52}NH_{12}$ | in-plane pyridinic, unsaturated | −3.39 | 1.29 | N | 2.11 |
| $C_{53}NH_{11}$ | pyrrolic | −1.26 | 1.15 | C | 2.40 |
| $C_{53}H_{11}NO$ | NO group | −2.43 | 1.06 | O | 1.82 |
| $C_{54}H_{11}NH_2$ | amino group | −1.37 | 1.22 | C | 2.36 |

It is obvious from the results presented in **Table 2** that $E_{ads}$(Al) is not directly linked to the charge transferred from Al to the substrate. However, this charge transfer is crucial for the conductivity of the studied GNRH – in their bare state (before Al adsorption) five of them are semiconductors with a small band-gap, up to 0.4 eV (**Fig. 3**, shaded areas; for whole-range DOS plots, see Supplementary Information, **Fig. S1**). This is expected, as it is known that graphene nanoribbons narrower than 10 nm are semiconductors [13]. The only exceptions are GNRH doped with an in-plane pyridinic N (both saturated and unsaturated, $C_{52}NH_{14}$ and $C_{52}NH_{12}$), which are conductive in their bare state. For the semiconducting systems, upon Al adsorption, the charge is transferred to the substrate, whose Fermi level shifts up (**Fig. 3**) and result in conductive Al@GNRH systems, with no band-gap. As shown in **Fig. 3**, p states of adsorbed Al are located mostly above Fermi level, *i.e.*, they are empty. Therefore, functionalization of GNRH with Al can be considered as a way to improve the material's conductivity, if needed for a given application. Again, the only exceptions are GNRHs doped with an in-plane pyridinic N, $C_{52}NH_{14}$ and $C_{52}NH_{12}$, for which we observe band gap opening upon Al adsorption. All investigated systems have total magnetization equal to zero, reflected in the symmetrical spin-up and spin-down states (**Fig. 3**). The only exception is bare GNRH doped with saturated in-plane pyridinic N ($C_{52}NH_{14}$, **Fig. 3**, shaded area), with total magnetization 0.87 $\mu_B$, but upon Al adsorption it drops to zero.

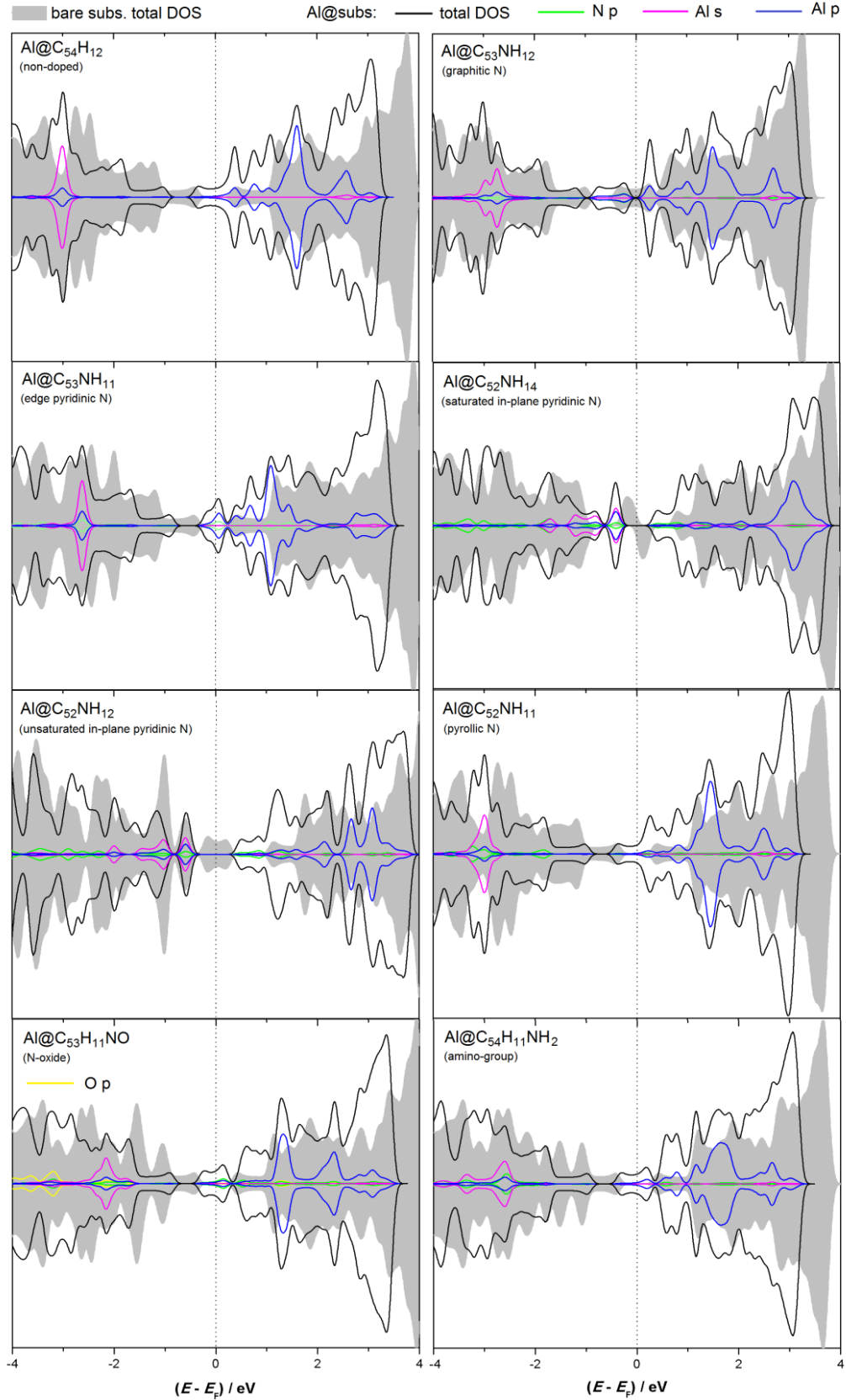

**Figure 3.** DOS plots for the studied GNRH models before and after Al adsorption on the preferential sites. For each model, the type of N-defect is stated. For the bare models, only total DOS is shown. For Al@substrate systems, Al s and p states and N p states are given as well. In the case of Al on N-oxide doped GNRH, O p states are also shown.

Since the strongest Al bonding was found for the cases of GNRH doped with in-plane pyridinic N (both saturated and not) and N-oxide, we look at them more closely. First, we focus on their charge difference plots (**Fig. 4**), calculated using Eq. (2). In all three cases, we observe that Al suffers both charge gain and charge loss – while some of its orbitals are emptied, others gain some (less) charge (**Fig. 4**). Looking at the cases of Al@$C_{52}NH_{14}$ and Al@$C_{52}NH_{12}$ (saturated and unsaturated forms of in-plane pyridinic N-doped GNRH, **Fig. 4 (a)** and **(b)**), we notice that the main areas of charge gain and charge loss are very similar for both structures. The charge is transferred from Al to N and its surrounding C atoms, where it is stored in the somewhat disrupted π cloud. In case of Al on GNRH with NO group the situation is somewhat similar. However, here the charge is transferred to both O and N, as well as $p_z$ orbitals of surrounding C atoms.

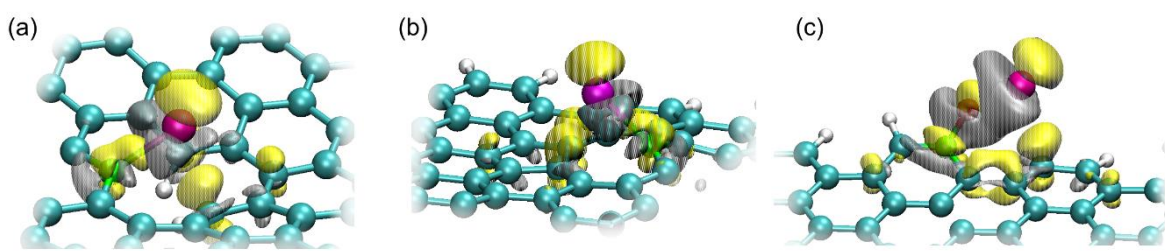

**Figure 4.** Charge difference plots for Al adsorbed on: (a) saturated in-plane pyridinic N-doped GNRH ($C_{52}NH_{14}$), (b) unsaturated in-plane pyridinic N-doped GNRH ($C_{52}NH_{12}$) and (c) N-oxide doped GNRH ($C_{53}H_{11}NO$). Isosurface values are ±0.003 eV Å$^{-1}$. Yellow isosurfaces indicate charge gain, while gray isosurfaces indicate charge loss. Graphical representations were made using VMD [33].

Next, we inspect the DOS plots for the three chosen systems more closely (**Fig. 5**). For the cases of pyridinic in-plane moiety (both saturated and unsaturated), Al is interacting directly with nitrogen atoms (**Fig. 5**, **a** and **b**, **Table 2**). Nitrogen s and p states are strongly hybridized and span over a wide energy window. In both cases, we see that Al s and $p_z$ states are hybridized and interact strongly with the sp$^2$ states of N. Al $p_x$ and $p_z$ states are dominantly empty and located above the Fermi level. This is because of the charge transfer from the Al adatom to the N-doped GNRH. In the case of N-oxide doped GNRH ($C_{53}H_{11}NO$) Al is interacting directly with the edge –NO group. Oxygen s and p states are strongly hybridized and largely filled due to the charge transfer from Al to O from the –NO group (**Fig. 5(c)**). The mode of interaction of Al with the –NO group is similar to that of pyridinic nitrogen. Namely, s and $p_z$ states of Al are strongly hybridized with a certain contribution of Al $p_x$ states.

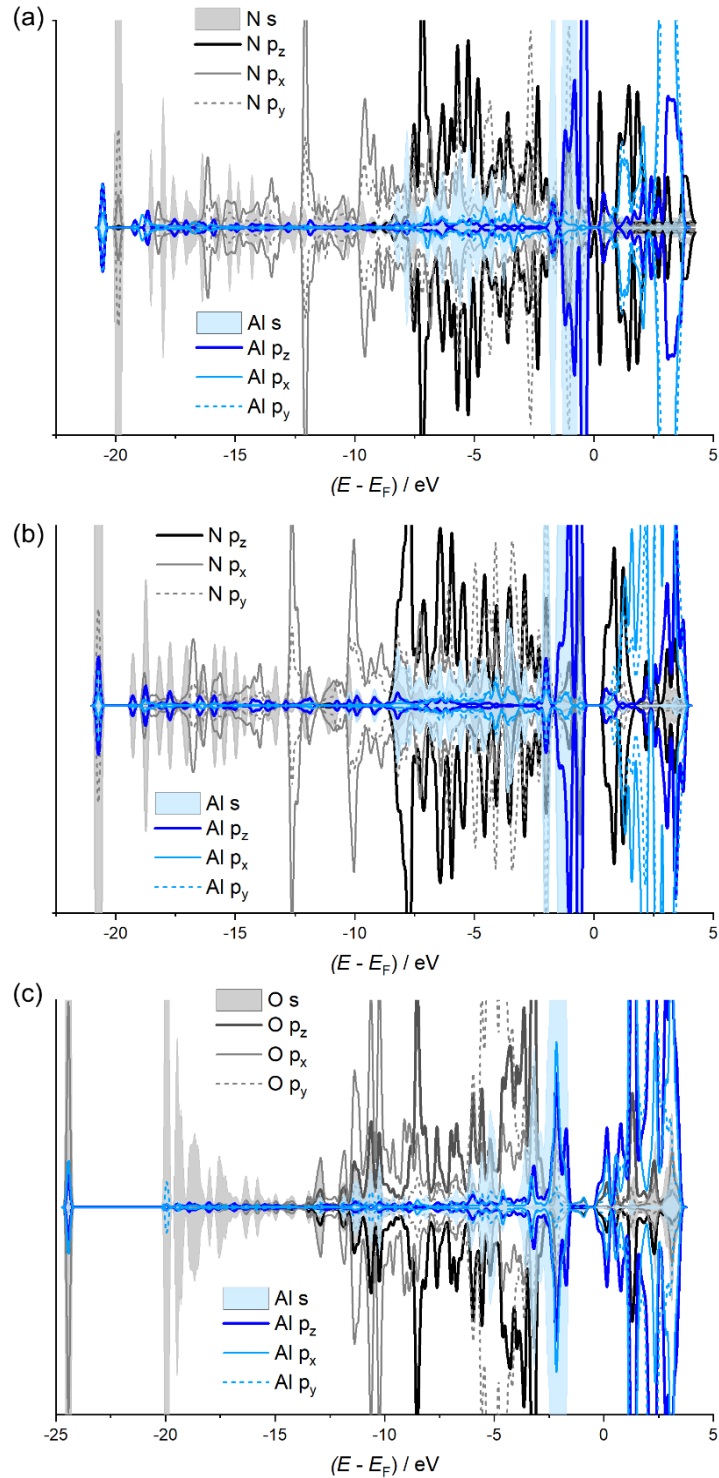

**Figure 5.** Detailed DOS plots for Al adsorbed on GNRH doped with: (a) saturated in-plane pyridinic N, (b) unsaturated in-plane pyridinic N and (c) N-oxide. All states of Al and the atom interacting directly with Al are shown (for GNRH doped with in-plane pyridinic N, both saturated and not, that is N, while for GNRH doped with NO that is O, as can be seen in **Fig. 2**).

As was mentioned in section 2, all of the results were obtained using dispersion-corrected PBE-D2 since it is known that this type of interaction can be significant when it comes to metals' interaction with graphene-based materials. However, in order to estimate the size of this effect, we have also calculated dispersion non-corrected Al adsorption energies on the chosen substrates. We find that for the studied systems Al adsorption energy calculated using PBE-D2, $E_{ads}$(Al), is a linear function (adj. $R^2$ = 0.996, Supplementary Information - **Fig. S2**) of Al's adsorption energy calculated without dispersion interaction ($E_{ads,nd}$(Al)), more precisely: $E_{ads}$(Al) = (1.023±0.006)×$E_{ads,nd}$(Al) − (0.16±0.01) eV. This finding is similar to previously reported relationships between $E_{ads}$ and $E_{ads,nd}$ for various adsorbates on graphene-based materials modelled using infinite graphene plane [5,11,34].

In light of the ongoing search for novel metal-ion batteries, we discuss the studied substrates as potential candidates for an electrode material in an Al-ion battery. As the simplest criterion for the possibility of reversible adsorption/desorption of Al during battery operation, we compare the reported values of $E_{ads}$(Al) to Al's cohesive energy ($E_{coh}$(Al) = 3.39 eV *per* atom [35]. Seven out of eight reported Al adsorption energies are smaller (absolute values) than $E_{coh}$(Al), indicating that Al phase separation could occur and hinder the battery operation. The only exception is GNRH doped with unsaturated in-plane pyridinic N, in which case Al's adsorption energy is equal to its cohesive energy. However, as we have mentioned before, such a system is highly unlikely to exist under standard conditions since the dangling bonds tend to saturate. In that sense, the next best potential candidate is GNRH doped with saturated in-plane pyridinic N. Another requirement for a good electrode material is good electric conductivity. As we have seen, most investigated substrates are small gap semiconductors, which turn to conductors upon Al adsorption, while GNRHs doped with in-plane pyridinic N act exactly in the opposite way. Even though the adsorption energy is more negative in these cases, the band gap opening upon Al bonding could be problematic for potential battery operation.

Since our results imply that the presence of O-containing moieties could benefit the nanoribbon's reactivity towards Al, it is worth mentioning that we have also investigated the possibility of Al adsorption on *p*-graphene functionalized with one O or OH group *per* $C_{54}$ (($3\sqrt{3}\times3\sqrt{3}$)$R$30°) supercell, which resulted in AlO or AlOH phase separation, *i.e.* the O/OH groups detachment from the graphene surface (Supplementary Information, **Fig. S3**). Evidently, the Al−O interaction was too strong for what we wanted to achieve in this case (possibility of reversible adsorption/desorption of Al). It is known that O-containing groups stabilize each other on the graphene surface [36]. However, the conductivity of the materials gets lower with rising O concentration, so once again (as so often in Materials Science), we are left to search for the optimal values of all the parameters required for a successful chosen application.

## 4. Conclusions

We performed DFT calculations to investigate the possibility of enhancing the adsorption of aluminium onto graphene. We have found that the simple introduction of edges, *i.e.* switching from ideal graphene to graphene nanoribbons, does not affect the energetics of Al adsorption significantly. Moreover, the influence of the adsorption site distance from the edge of the non-doped GNRH on Al's adsorption energy is negligible. N-doping of graphene nanoribbons affects the nanoribbon's reactivity towards Al, but the outcome strongly depends on the type of N-containing defect. While graphitic N induces minimal changes in this regard, we find that the introduction of the NO group at the edge results in Al adsorption nearly twice as strong as on pristine graphene and doping with pyridinic N in even stronger binding by about 0.3 eV. In all the investigated cases, there is a significant charge transfer from Al to the substrate. That way, the substrates, which were semiconductors with a small band-gap in the bare state, receive enough charge to become conductive, with in-plane pyridinic N doped ribbon as the only exception. In addition, the functionalization of semiconducting graphene nanoribbons with Al can be considered as a way for improving the material's conductivity. The charge transferred from Al to the N-doped GNRH is not located only at the site where the interaction occurs (usually the N site), but it is also partially stored in the graphene plane around the dopant site. Overall, selective doping and functionalization of graphene nanoribbons with N-containing moieties positively affect Al storage and could lead to the development of advanced materials for the next generation of multivalent ion rechargeable batteries.

## Acknowledgments

This research was supported by the Serbian Ministry of Education, Science and Technological Development (Contract number: 451-03-68/2020-14/200146). This research was sponsored in part by the NATO Science for Peace and Security Programme under grant no. G5729. N.V.S. acknowledges the support provided by the Swedish Research Council through the project no. 2014-5993. We also acknowledge the support from the Carl Tryggers Foundation for Scientific Research (grant no. 18:177), Sweden. The computations and data handling were enabled by resources provided by the Swedish National Infrastructure for Computing (SNIC) at NSC center of Linköping University, partially funded by the Swedish Research Council through grant agreement no. 2018-05973.

# SUPPLEMENTARY INFORMATION

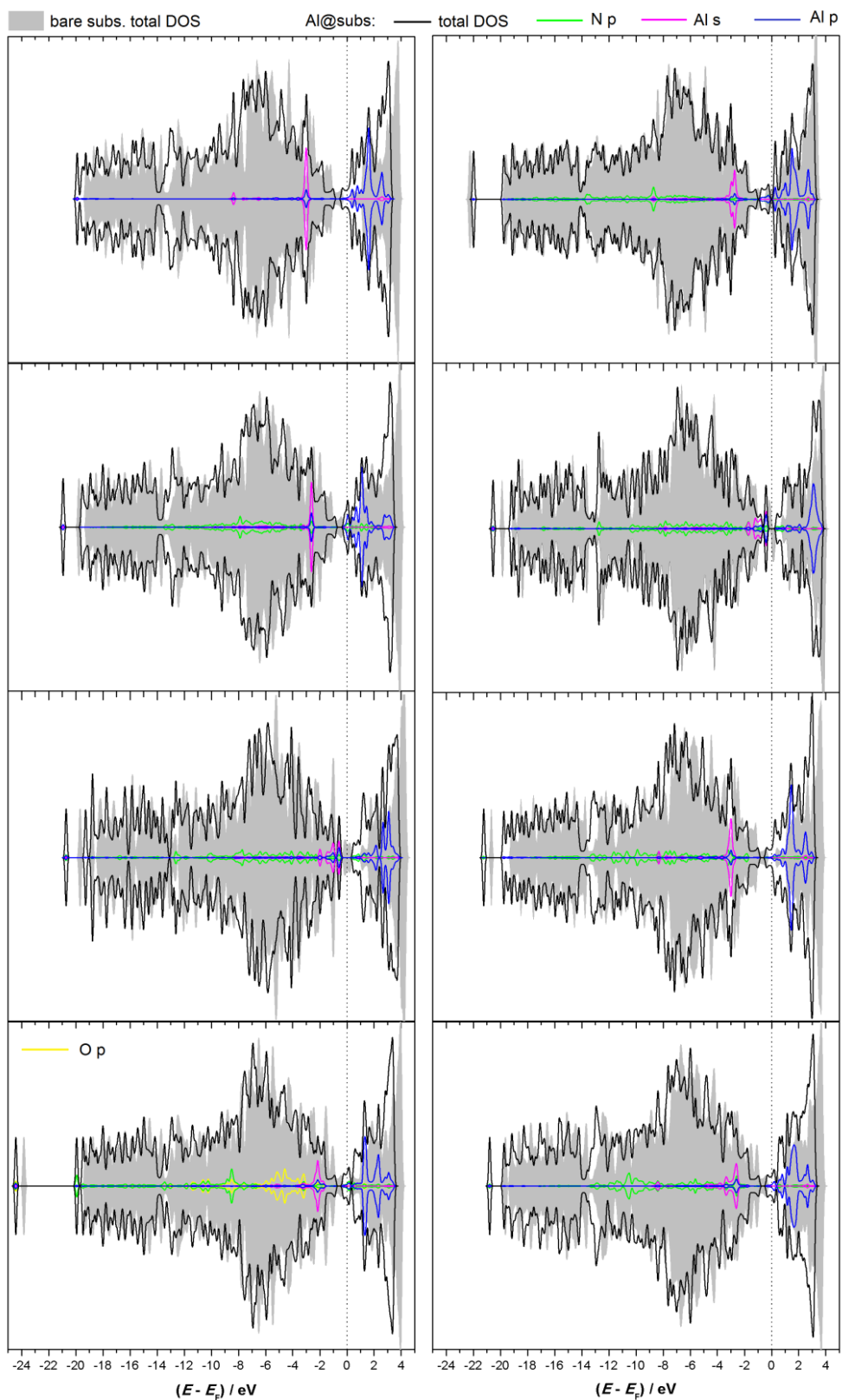

**Figure S1.** Whole-range DOS plots for the studied GNRH models before and after Al adsorption on the preferential sites. For the bare models only total DOS is shown. For Al@substrate systems Al s and p states, as well as N p states are given, while for the case of N-oxide, where Al interacts directly O, oxygen's p states are also shown.

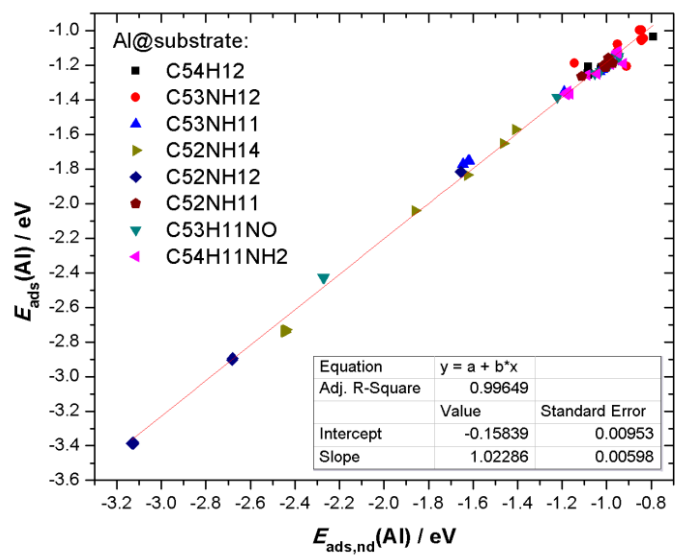

**Figure S2.** The relationship between Al adsorption energies on the studied substrates, calculated using dispersion non-corrected PBE ($E_{\text{ads,nd}}(\text{Al})$), and dispersion corrected PBE-D2 ($E_{\text{ads}}(\text{Al})$).

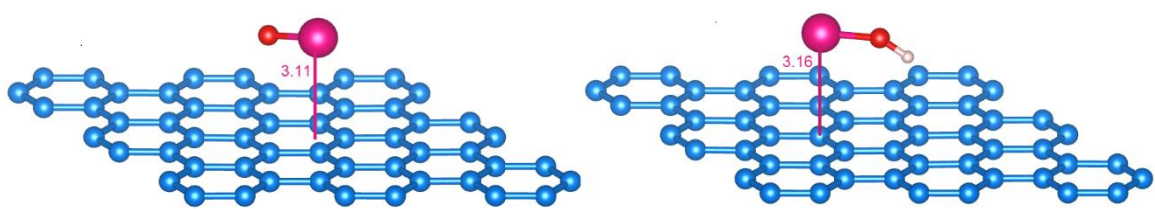

**Figure S3.** Optimized structures of oxidized graphene upon Al adsorption. Left: O-graphene ($C_{54}O$ model), right: OH-graphene ($C_{54}OH$ model). The distance from Al to the graphene basal plane is given in Ångströms. Graphical representations were made using VESTA [30].